\documentclass[prb,showpacs,twocolumn,floatfix]{revtex4}
\usepackage{graphicx}
\usepackage{bm}
\usepackage{dcolumn}

\begin{document}


\title{
 The free surface of superfluid $^{\bm 4}$He at zero temperature
}
\author{J. M. Mar\'\i n, J. Boronat, and J. Casulleras}

\affiliation{Departament de F\'\i sica i Enginyeria Nuclear,
  Universitat Polit\`ecnica de Catalunya,
  Campus Nord B4-B5, E-08034 Barcelona, Spain}

\date{\today}


\begin{abstract}
The structure and energetics of the free surface of superfluid $^4$He are
studied using the diffusion Monte Carlo method.  Extending a previous
calculation by Vall\'es and Schmidt, which used the Green's function Monte
Carlo method,  we study the surface of liquid $^4$He within a slab
geometry using a larger number of particles in the slab and an updated
interatomic potential.  The surface tension is accurately estimated from
the energy of slabs  of increasing surface density and its value is close
to one of the two existing experimental values. Results for the density
profiles allow for the calculation of the surface width which shows an
overall agreement with recent experimental data. The dependence on the
transverse direction to the surface of other properties such as the
two-body radial distribution function, structure factor, and one-body
density matrix is also studied. The condensate fraction, extracted from
the asymptotic behavior of the one-body density matrix, shows an
unambiguous enhancement when approaching the surface.

\end{abstract}


\maketitle


\section{Introduction}

The interest of quantum many-body theory in inhomogeneous superfluid
$^4$He has been active for many years and continuously enriched by the
achievement of new experimental realizations of confining
geometries.~\cite{book} The most classical, and consequently best known,
correspond to $^4$He films adsorbed on different
substrates~\cite{kro1,kro2} and to $^4$He clusters produced by free jet
expansions through very thin nozzles.~\cite{special} More recently, and
still with some open questions, interest has been devoted to $^4$He in
porous media such as vycor or aerogel~\cite{chan,zassen,glyde1} and $^4$He
adsorbed in carbon nanotube
bundles~\cite{teizer,lasja,cole,gordillo,leandra} forming quasi-one
dimensional structures. In common with other fluids currently under
study,  the effects of a confined geometry upon its microscopic properties
are most interesting. Additionally, liquid $^4$He poses an unavoidable
quantum behavior which manifests in its own existence as a liquid even at
zero temperature, with a superfluid phase and a partially occupied
Bose-Einstein state. Moreover, its extreme purity allows for a much
cleaner extraction of data which is usually impossible with other
liquids. 

One of the fundamental features raised by the inhomogeneous situations 
described above is the emergence of the free surface of superfluid
$^4$He.~\cite{edwards} For a long time, the main concern has been to
understand how distinctive properties of this quantum liquid, like its
superfluid density and condensate fraction, behave when the density goes
to zero through a free surface.~\cite{campbell} Also the surface tension,
the form of its density profile, the value of the surface width, and the
role of the surface excitations in its dynamics have been theoretically
and experimentally studied.~\cite{kro2,gernoth,dalfovo,toennies,barranco}
The most direct information on the density profile and interfacial width
has been obtained by x-ray reflectivity~\cite{lurio1,lurio2,penanen} and
ellipsometric~\cite{osborne} measurements. The latter was carried out by
Osborne,~\cite{osborne} who assuming a Fermi function for the profile,
extracted a 10-90 \% width of 9.4 \AA\ at 1.8 K. The most recent data
correspond to the x-ray measurement of Penanen \textit{et
al.}~\cite{penanen} who reported a value 6.5 \AA\ at 0.45 K, a lower
temperature than the first measure performed by Lurio  \textit{et
al.}~\cite{lurio2}    at 1.13 K who reported a wider surface of 9.1 \AA.
The relative spread of the experimental results on the microscopic
characteristics of the density profile points out the difficulties which
experimental setups face to carry out a clean extraction of the data and
its reduction to the ground state, i.e., to zero temperature.

The difficulty of the experimental work on the study of the $^4$He free
surface is also reflected in the long-time effort to measure accurately the
surface tension. Since the first work by Urk \textit{et al.}~\cite{urk} in 1925, the
number of groups and different techniques used for this measure is
unusually large. Recent data starts with Iino \textit{et
al.}~\cite{iino} who
measured the surface tension using the surface-wave resonance method
obtaining an estimation at zero temperature of $\sigma=0.257$ K\AA$^{-2}$.
Later on, Roche \textit{et al.}~\cite{roche} obtained a slightly larger value  
$\sigma=0.274$ K\AA$^{-2}$ from the frequencies of capillary waves with
wavelength in the micron range. In this second work, the difference in the
values obtained for $\sigma$ is imputed to a possible inaccuracy in the
work of  Iino \textit{et al.}~\cite{iino} related to the treatment of the meniscus at
the edge of the box. However, this argument was refuted posteriorly by Nakanishi and
Suzuki~\cite{suzuki} who confirmed the result obtained in Ref.
\onlinecite{iino}. Recently, Vicente
\textit{et al.}~\cite{vicente} performed a new measure of the surface tension using the
vibration modes of levitated $^4$He drops, and the result was in perfect
agreement with the larger value obtained by Roche \textit{et
al.}.~\cite{roche} 

The surface tension, surface width, and density profile of the free $^4$He
surface have been calculated using both density
functional~\cite{dalfovo,guirao,szyb} and microscopic 
theory.~\cite{gernoth,chin,pieper,valles,ceper} Density functional theory relies on functional forms of the energy
which depend on the one-body density and some parameters which are adjusted
to reproduce selected experimental data. These functionals have been
progressively improved by the inclusion of non-local contributions and specific 
terms to reproduce
the experimental static response function.~\cite{dalfovo} Contrarily to the first models,
in which the surface tension was considered an input to fix parameters, the
most modern ones are able to give predictions for $\sigma$. The results for the
surface tension are in an overall agreement with the experimental data from
Ref. \onlinecite{roche},
and the surface width is $\sim 6$ \AA. The methodology of microscopic
approaches is different since the starting point in all of them is the 
Hamiltonian of the liquid containing realistic interparticle interactions.
In this respect, the variational method has been specially useful due to the  
hard-core of the He-He interaction. In fact, the most extensive work in the
field of inhomogeneous $^4$He has been carried out in the variational
framework of the hypernetted chain (HNC) and correlated basis function
(CBF) theories. Of special relevance in this field is the continued work by
Krotscheck and co-workers,~\cite{krobook} who have carried out a nearly exhaustive study of
$^4$He in reduced geometries both for the ground and the excited states.

Quantum Monte Carlo methods~\cite{hammond} have also been applied to the study of the free
surface of liquid $^4$He. Using a slab geometry, Vall\'es and
Schmidt~\cite{valles}
calculated the ground-state properties of the surface by means of the
variational Monte Carlo (VMC) method with a trial wave function containing
two- and three-body correlations. They extracted a surface tension 
$\sigma=0.272$ K\AA$^{-2}$ (again very close to data from Ref. \onlinecite{roche}) 
from an analytical fit to the dependence of the energy with the inverse of the 
surface density. Galli and Reatto~\cite{shadow} studied a $^4$He slab using a
shadow wavefunction and the VMC method; they obtained  $\sigma=0.31(1)$ K\AA$^{-2}$  and a
surface thickness $\sim 5$ \AA.  The variational constraints can be removed
by applying the essentially exact diffusion Monte Carlo (DMC)
and Green's function Monte Carlo (GFMC) methods. At present, the only
application of these methods to a slab corresponds to a GFMC calculation by
Vall\'es and Schmidt~\cite{valles} who estimated a surface tension of 0.265 K\AA$^{-2}$
and a small value for the width ($\sim 4$ \AA) probably due to the small
number of particles used in the simulation. Recently, Draeger and
Ceperley~\cite{ceper}
carried out a similar study but at finite temperature using the path
integral Monte Carlo (PIMC) method. This work does not contain results for
$\sigma$ and surface width since its main motivation was the study of the
enhancement of the Bose-Einstein condensate fraction at the surface.

In the last years, a renewed interest in the study of the $^4$He surface has
emerged since the first theoretical observation by Lewart \textit{et
al.}~\cite{lewart}
of a large increase of the condensed fraction near the surface of $^4$He
drops. The relevance of this result in the long way for searching a
Bose-Einstein condensed state was put forward by Griffin and
Stringari~\cite{stringari} who
studied the low density surface using a generalized Gross-Pitaevskii
equation. A subsequent work by Galli and Reatto,~\cite{shadow} using the 
VMC method with a trial wave function based on the shadow model,
pointed out that the
introduction of fluctuations on the surface, mainly the zero point motion
of ripplons, can reduce significantly the enhancement of $n_0$. Their results
show an increase of $n_0$ up to a maximum value $0.5$ and a decrease
to zero when the density $\rho(r)$ approaches zero. In order to shed light
on these two different predictions, Draeger and Ceperley~\cite{ceper} calculated $n_0$
in a $^4$He slab using PIMC. Their results show also a clear enhancement of
the condensate fraction up to values larger than $0.9$ and a small decrease
when approaching the outer part of the surface. 
Therefore, fluctuations can
arise and reduce $n_0$ but their effects could be significantly smaller than 
in the VMC calculation of Ref. \onlinecite{shadow}. 
A conclusive statement would require from a direct
experimental measure which, for example, in the bulk has not yet been
possible. Nevertheless, at present there are two experiments that seem to
confirm the increase of $n_0$. The first one was derived from a quantum
evaporation experiment,~\cite{wyatt} and the second and more recent, form
deep-inelastic neutron-scattering on thick layers of $^4$He on an MgO
substrate.~\cite{pearce}

In the present paper, we present a DMC calculation of the ground-state
properties of a free $^4$He surface at zero temperature. To some extent,
our work represents an update of the GFMC calculation of Vall\'es and
Schmidt~\cite{valles} in the sense of considering a more accurate interatomic potential
and a larger number of particles in the simulation to achieve thicker slabs and therefore
a more realistic study of the surface. Moreover, we have included in the
present study topics like the density profile form, two-body distribution
functions and structure
factors, and mainly the study of the one-body density matrix which were not
analyzed in Ref. \onlinecite{valles}. The exact character of the DMC method, within the
statistical noise, allows for an accurate study of all the surface
properties without the bounds imposed by a variational treatment. The
methodology used in the analysis is, regarding the main inputs, the same that we
have used previously in the study of bulk liquid $^4$He and $^3$He, and that
has shown an overall agreement between theory and
experiment.~\cite{borobook}

Our paper is organized as follows. In the next section, we review the DMC
method which works with a second-order approximation for the short-time
Green function. The specific details of the simulation for the slab
geometry are also discussed, with special emphasis on the requirements for
the trial wave function used for importance sampling. The results obtained,
for both structure and energetics of the slab, are presented in Section III.
Finally, Section IV comprises additional discussion on the results and the
main conclusions.

\section{Diffusion Monte Carlo and the slab geometry}

The DMC method is nowadays a well established tool for solving the
many-body imaginary-time Schr\"{o}dinger equation,
\begin{equation}
- \frac{\partial \Psi({\bf R},t)}{\partial t} = (H-E)\, \Psi({\bf R},t) \ ,
\label{dmc.eq1}
\end{equation}
where ${\bf R} \equiv ({\bf r}_1,\ldots,{\bf r}_N)$ is a $3N$-dimensional
vector (\textit{walker}) and $t$ is the imaginary time measured in
units of $\hbar$. The time-dependent wave function of the system 
$\Psi({\bf R},t)$ can be expanded in terms of a
complete set of eigenfunctions $\phi_i({\bf R})$ of the Hamiltonian,
\begin{equation}
\Psi({\bf R},t)=\sum_{n}c_n \, \exp \left[\, -(E_i-E)t \, \right]\,
\phi_i({\bf R})\ ,
\label{dmc.eq1b}
\end{equation}
where $E_i$ is the eigenvalue associated to $\phi_i({\bf R})$.  The
asymptotic solution of Eq.~(\ref{dmc.eq1}), for any value $E$ close to
the energy of the ground state and for long times ($t \rightarrow
\infty$), gives $\phi_0({\bf R})$, provided that there is a nonzero
overlap between $\Psi({\bf R},t=0)$ and the ground-state wave function
$\phi_0({\bf R})$.

An efficient solution of Eq.~(\ref{dmc.eq1}) requires from the use of
importance sampling. It is introduced by solving the Schr\"{o}dinger
equation for  the wave function
\begin{equation}
f({\bf R},t)\equiv \psi({\bf R})\,\Psi({\bf R},t)\ ,
\label{dmc.eq2}
\end{equation}
$\psi({\bf R})$ being a time-independent trial wave function able to
describe the system at a variational level. 
Introducing the Hamiltonian of the system
\begin{equation}
H=-\frac{\hbar^2}{2\,m} \, \bm{\nabla}^2_{{\bf R}} 
+ V({\bf R})\ ,
\label{dmc.eq3}
\end{equation}
Eq.~(\ref{dmc.eq1}) turns out to be
\begin{eqnarray}
-\frac{\partial f({\bf R},t)}{\partial t}  & = &  -D\, 
\bm{\nabla}^2_{{\bf R}}
f({\bf R},t)+D\, \bm{\nabla}_{{\bf R}} \left( {\bf F}({\bf R})
\,f({\bf R},t)\,
\right) \nonumber \\
 & & +\left(E_L({\bf R})-E \right)\,f({\bf R},t)  \ , \label{dmc.eq4} 
\end{eqnarray}
with $D=\hbar^2 /(2m)$, $E_L({\bf R})=\psi({\bf R})^{-1} H \psi({\bf R})$
the local energy, and
\begin{equation}
{\bf F}({\bf R}) = 2\, \psi({\bf R})^{-1} 
\bm {\nabla}_{{\bf R}} \psi({\bf R})
\label{dmc.eq5}
\end{equation}
is the drift force which guides the diffusion process.

The r.h.s. of Eq.~(\ref{dmc.eq4}) may be written as the action of three
operators $A_i$ acting on the wave function $f({\bf R},t)$,
\begin{equation}
-\frac{\partial f({\bf R},t)}{\partial t} = (A_1+A_2+A_3)\, 
f({\bf R},t) \equiv A\, f({\bf R},t)
\label{dmc.eq4p}
\end{equation}
The first one, $A_1$, corresponds to a free
diffusion with a a diffusion coefficient $D$; $A_2$ acts as a driving force
due to an external potential, and  finally $A_3$ looks like a birth/death
term. Equation (\ref{dmc.eq4p}) is transformed to the integral form
\begin{equation}
     f({\bf R}^{\prime},t+\Delta t) =\int G({\bf R}^{\prime},{\bf R},
\Delta t)\, f({\bf R},t)\, d{\bf R} \ ,
\label{dmc.eq6}
\end{equation}
by introducing the Green function 
\begin{equation}
    G({\bf R}^{\prime},{\bf R}, \Delta t) =  
    \left \langle\,
{\bf R}^{\prime}\, | \, \exp(-A \Delta t)\, |\, {\bf R}\, \right \rangle.
\label{dmc.eq7}
\end{equation}
with $A \equiv A_1 + A_2 + A_3$.

In the DMC method, Eq.~(\ref{dmc.eq7}) is iterated repeatedly until to reach 
the asymptotic regime
$f({\bf R},t \rightarrow \infty)$, a limit in which one is effectively
sampling the ground state. In our implementation of the method we use 
a second-order expansion (quadratic DMC),~\cite{casu}
\begin{eqnarray}
 \lefteqn{\exp \left( -A \Delta t \right)  = }  \label{dmc.eq8} \\ 
   &  &   \exp \left( -A_3 \frac{\Delta
t}{2} \right ) \, \exp \left( -A_2 \frac{\Delta t}{2} \right ) \,
\exp \left( -A_1 \Delta t \right )  \nonumber  \\
&  & \times \ \exp \left( -A_2 \frac{\Delta t}{2} \right ) \,
\exp \left( -A_3 \frac{\Delta t}{2} \right )  + {\cal O} \left( (\Delta t)^3
\right)
\, , \nonumber
\end{eqnarray}
which has proved to be a good compromise between algorithmic complexity 
and efficiency.

The study of the free surface is made by simulating a slab which grows
symmetrically in the $z$ direction and with periodic boundary conditions 
in the $x-y$ plane. The slab geometry is probably the best suited one to
analyze the free surface since it does not present neither the substrate influence
of films  nor the curvature effects of drops. It is also true that the
existence of two surfaces can introduce some residual influence
between them. However, we have checked in our simulations that this effect,
at least for the larger slabs, is completely negligible. 

The implementation of DMC requires of a model for the trial wave function
$\psi({\bf R})$ with the basic physical features of the system under study.
In the present case, one \textit{a priori} knows that $\psi({\bf R})$ 
becomes zero in two situations: when two atoms get closer than the core of 
its interaction, and when an atom tries to escape from the surface. Notice
that at zero temperature no vapor is present out of the liquid. Having in
mind these arguments, we consider
\begin{equation}
\psi({\bf R})  = \psi_{\text J}({\bf R}) \ \phi({\bf R}) \ ,
\label{trial1} 
\end{equation} 
with a Jastrow correlation factor accounting for  dynamical correlations
\begin{equation}
\psi_{\text J}({\bf R}) = \prod_{i<j}^N f(r_{ij})  \ ,
\label{trial2}
\end{equation}
and a confining term, factorized in the form
\begin{equation}
\phi({\bf R})= \prod_{i=1}^N h(z_i)  \ .   
\label{trial3}
\end{equation}
The two-body correlation function is the same we used in the study of the
bulk liquid.~\cite{casu} It was proposed by Reatto~\cite{reatto} and incorporates, in an approximate
way, the medium range behavior of $f(r)$ observed in the Euler-Lagrange
functional optimization,
\begin{equation}
f(r) = \exp \left[ - \frac{1}{2} \, \left( \frac{b}{r} \right)^5
-\frac{L}{2} \, \exp \left[ - \left( \frac{r - \lambda}{\Lambda} \right)^2
\right] \right] \ .
\label{trial4}
\end{equation}
The confining function is chosen of Fermi type, like in Refs.
\onlinecite{valles} and \onlinecite{liu},
\begin{equation}
h(z)= \left\{ 1 + \exp [ \, k ( \, |z-z_{\text{cm}}| - z_0)  ] \right\}^{-1}   \ ,
\label{trial5}
\end{equation}
with parameters $k$ and $z_0$ controlling the width and location of the
interface, respectively. Possible and spurious kinetic energy contributions due to the
movement of the center of mass are removed by subtracting $z_{\text{cm}}$
to each particle coordinate $z$ in Eq. (\ref{trial5}).

\section{Results}

The properties of the free surface of liquid $^4$He have been studied by
performing DMC calculations with an increasing number of particles $N$=54,
108, 162, 216, 324, and a fixed $x-y$ area of $A=290.3$ \AA$^2$. The
interatomic potential is the HFD-B(HE) model proposed by Aziz \textit{et
al.}~\cite{aziz} that has proven its high accuracy in previous DMC calculations of the
density dependence of the pressure and speed of sound in bulk $^4$He and
$^3$He liquids.~\cite{borobook} As a matter of comparison with the GFMC calculation of
Vall\'es and Schmidt,~\cite{valles} and in order to establish
the influence of the potential in the
surface properties, we have calculated the energy of the system and the
surface tension using the HFDHE2 model, also proposed by Aziz and
collaborators.~\cite{azizvell} 

In order to take into account size effects, due to the finite value of the
simulation area $A$ and periodic boundary conditions in the $x-y$ plane, we
have added to the energy tail corrections which are externally
calculated assuming that the particles are uncorrelated for distances in
this plane larger than half of the simulation square. The analytical
expressions for the tails are reported in Ref. \onlinecite{valles}. Applying this analysis,
we have verified that the energy per particle when doubling simultaneously
$A$ and $N$ does not change within the current statistical noise. Possible
bias in the DMC calculations coming from the time step and the mean population 
of walkers are also under control working, in both cases, with values which
are well in the asymptotic regime.

A final key point of a DMC simulation concerns the values of the
parameters in the trial wave function (\ref{trial4},\ref{trial5}).
This has been dealt
by performing a series of VMC calculations and searching the optimal set.
As expected, the
only parameter which clearly shows a dependence on the number of particles,
i.e., the size of the slab, is $z_0$; the particular values are reported in Table I.
The rest of the set has shown negligible dependence on $N$; their values
are $b=3.067$ \AA, $L=0.2$, $\lambda=5.112$ \AA, $\Lambda=1.534$ \AA, and
$k=1$ \AA$^{-1}$.

\subsection{Energies and surface tension}

\begin{table}
\centering
\begin{ruledtabular}
\begin{tabular}{ccdddd}
$N$  & $n_{\text c}$ (\AA$^{-2}$)  &  \multicolumn{1}{c}{$z_0$ (\AA)}  &  
\multicolumn{1}{r}{$E/N$ (K)}  & \multicolumn{1}{r}{$T/N$ (K)}  & 
\multicolumn{1}{c}{$\ \ \ \ \ \ \ E^I/N$ (K)}
\\  \hline
54   & 0.1860  &  3.60    &  -4.519(6) &  7.191(13)  &  -4.403(5)[-4.65]  \\
80   & 0.2756  &  5.41    &  -5.287(6) &  8.562(15)  &  -5.155(4)   \\ 
108  & 0.3720  &  7.30    &  -5.763(8) &  9.515(32)  &  -5.636(6)[-5.69] \\ 
162  & 0.5580  &  10.95   &  -6.261(7) &  10.950(34) &  -6.129(6)   \\
216  & 0.7441  &  14.60   &  -6.528(6) &  11.720(34) &  -6.390(7)   \\
324  & 1.1161  &  22.10   &  -6.766(11)&  13.075(40) &  -6.637(10)  \\
    & $\infty$ &          &  -7.267(13)&  14.32(5)   &  -7.121(10)
\end{tabular}
\end{ruledtabular}
\caption{Total ($E/N$) and kinetic ($T/N$) energies per particle as a function of the number
of particles $N$ and the coverage density $n_{\text c}$. $E^I/N$ is
calculated with the old Aziz potential (HFDHE2);~\protect\cite{azizvell} the results in squared
parenthesis are taken from Ref. \protect\onlinecite{valles}. Figures in parenthesis are the
statistical errors. $z_0$ is a variational parameter
present in the correlation factor $h(z)$ (\protect\ref{trial5}). The bulk
result is taken from Ref. \protect\onlinecite{casu} }
\end{table}

The DMC results of the energy per particle as a function of the number of
particles included in the simulation, with a fixed area $A$, is reported in
Table I. The coverage densities $n_{\text c}$, defined as
\begin{equation}
n_{\text c} = \frac{N}{A} = \int_{-\infty}^{\infty} dz \, \rho(z) \ ,
\label{coverage}
\end{equation}
with $\rho(z)$ the density profile, are also contained in the Table. When
$n_{\text c}$ increases, the energies approach the bulk value~\cite{casu} which appears
in the last row to help in the comparison. In the last column, we have
included results for the energy using the old Aziz (HFDHE2) potential. As
it is known from previous DMC calculations for the bulk,~\cite{casu} the energies
obtained with the HFDHE2 potential are slightly smaller (in absolute
value), the difference with the  HFD-B(HE) model being mainly due to the
increase in the potential energy. Results for the kinetic energy, which are
the same for the two potentials within error bars, are included in the Table showing
convergence to the bulk value (14.32 K).~\cite{pures} Enclosed in squared brackets in
the Table there are GFMC results from Ref. \onlinecite{valles} calculated at the same
coverages; the agreement with our results is better for the larger coverage
case but the lack of data at larger  $n_{\text c}$ leaves the comparison rather
incomplete in the most interesting region, i.e., for the largest slabs.

\begin{figure}
\centering
        \includegraphics[width=0.8\linewidth]{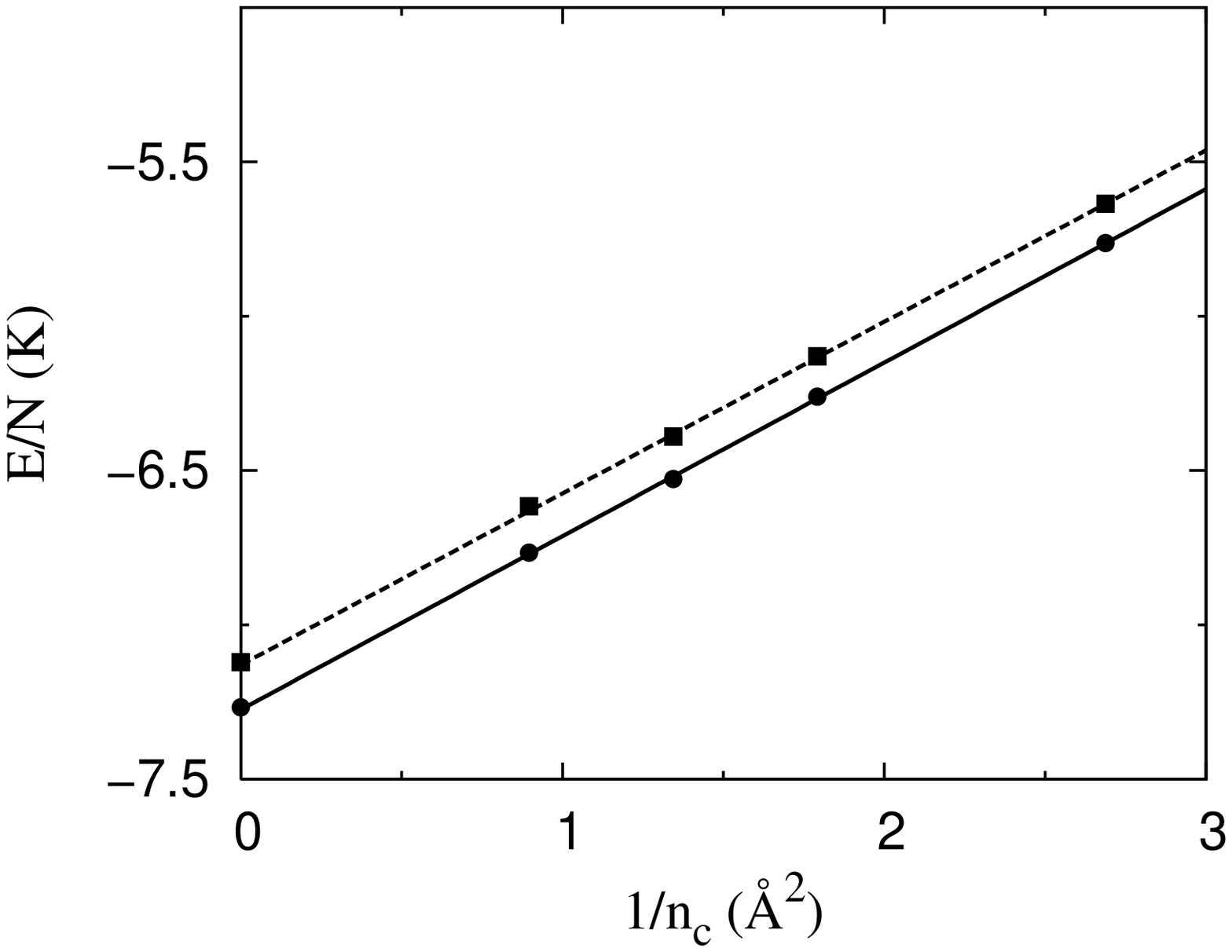}%
	\vspace{0.1cm}%
	\\
	\includegraphics[width=0.8\linewidth]{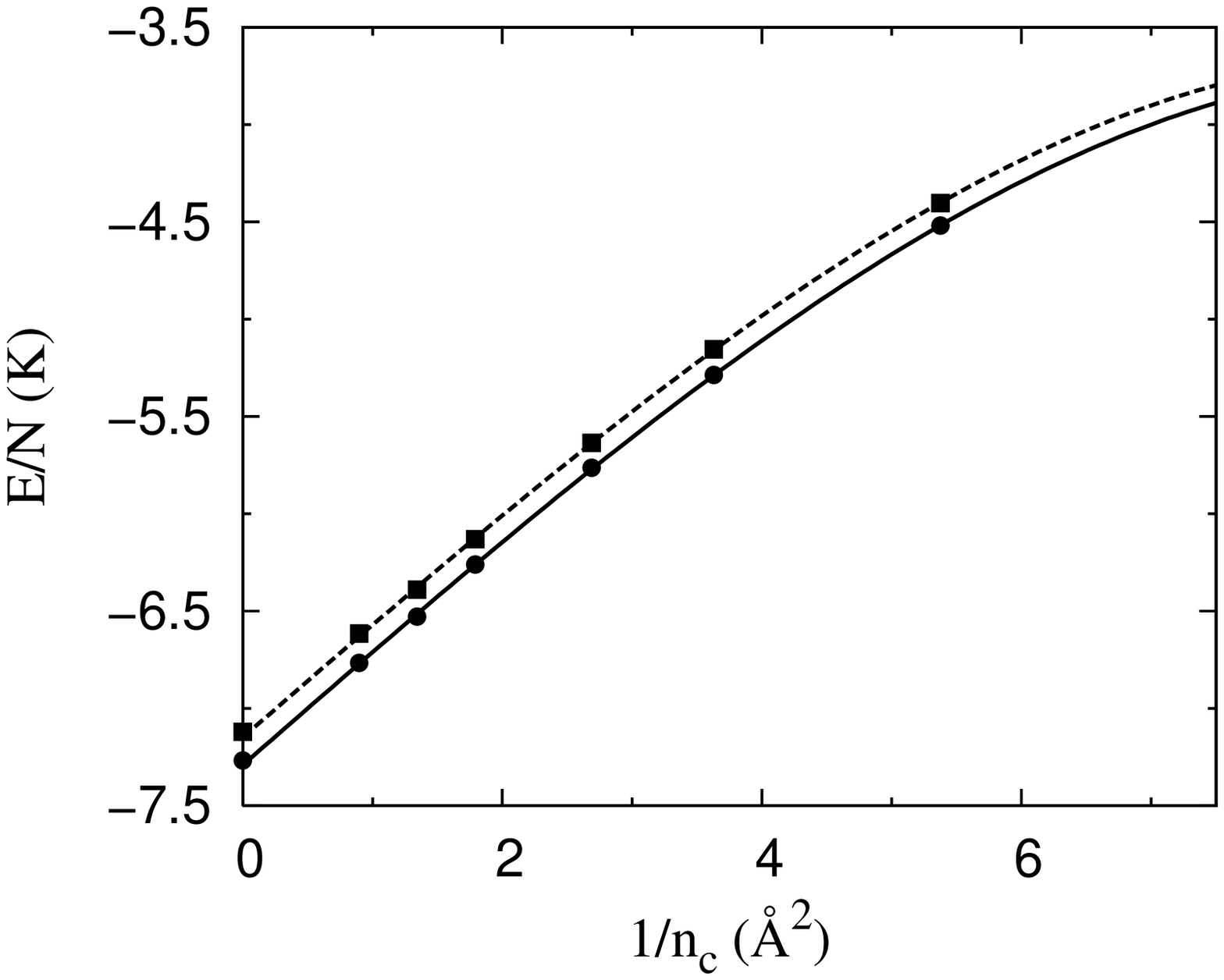}%
\caption{Energies per particle as a function of the inverse of the surface
coverage. Circles and squares stand for the calculations with the HFD-B(HE)
and the HFDHE2 potentials, respectively. \textit{Top}: Large $n_{\text
c}$ regime; solid and dashed lines correspond to numerical fits
(\protect\ref{expansion}) with 
$\gamma=0$. \textit{Bottom}: All the coverages; the lines are numerical
fits (\protect\ref{expansion})to the DMC data with $\gamma \neq 0$.}
\end{figure}

The energies per particle, as a function of the inverse of the surface
coverage ($1/n_{\text c}$), are shown in Fig. 1. 
As a limiting case ($n_{\text c}=\infty$), 
the energies corresponding to the bulk for the two potentials are also
plotted. Under general assumptions it is expected that the energy per
particle in the slab follows a polynomial expansion in terms of $1/n_{\text
c}$,~\cite{szyb}
\begin{equation}
\left(\frac{E}{N} \right) = \left(\frac{E}{N} \right)_\infty + \frac{2
\sigma}{n_{\text c}} + \frac{\gamma}{n_{\text c}^3} +
{\cal O}\left(\frac{1}{n_{\text c}^4} \right) \ .
\label{expansion}
\end{equation} 
In this expansion, $\sigma$ is the surface tension and the parameter $\gamma$ has been
related to the long-range interaction term of an hypothetical $^4$He
substrate.~\cite{szyb} In Fig. 1 (top), the energies for slabs with 
different
number of atoms are shown for values $1/n_c < 3$ \AA$^2$. In this regime,
corresponding to the largest slabs, the energy is a linear function of
$1/n_c$. A linear fit to the data, for both interatomic potentials, 
gives  $\chi^2/\nu =1-1.5$. From the linear coefficient we estimate the
surface tension; for the most accurate HFD-B(HE) potential
$\sigma=0.281(3)$ K\AA$^{-2}$, and  $\sigma=0.278(3)$ K\AA$^{-2}$ for the
old Aziz potential. Within the statistical errors, the surface tension is
the same for both potentials implying that changing the potential
introduces a shift in the energy which is to a large extent independent 
of the number of atoms. In the fit, the DMC energy of the bulk phase,
corresponding to $1/n_c=0$, has been included. However, it turns out
that the inclusion or not of the bulk energy in the fit does not modify the
result for the surface tension, and that the extrapolated energy to $1/n_c=0$
matches the bulk result, a feature that gives us additional confidence on
the accuracy of the slab energies.

In order to describe the whole data set it is necessary to introduce in the
fit higher-order terms (\ref{expansion}). 
To this end, we have introduced
a next-order term with $1/n_c^2$ or $1/n_c^3$, and tried to discriminate
which of the two options is more favored by our data. Using the third-power, 
as suggested by Szybisz~\cite{szyb} (\ref{expansion}), the resulting value 
for $\chi^2/\nu$ is 2.5, a value significantly smaller than the one
achieved with a quadratic law (8.0).
Therefore, and in spite of not being
completely conclusive, our data support Szybisz's analysis. The
corresponding fits are shown together with the DMC data in Fig. 1 (bottom).
The surface tension from this fit is $\sigma=0.291(4)$ K\AA$^{-2}$, nearly
compatible with our best prediction from the linear fit to the largest
$n_c$ values. On the other hand, $\gamma=-0.0023(2)$ K\AA$^{-6}$ roughly
half the approximated value obtained by Szybisz. 
     
The present result for the surface tension is a bit larger than the one
obtained by Vall\'es and Schmidt in a GFMC calculation with the old Aziz
potential (HFDHE2), $\sigma=0.265(6)$ K\AA$^{-2}$.~\cite{valles} 
That difference 
can be attributed to the fact that we have available data corresponding to 
large $n_c$ values, which is crucial for the estimation of $\sigma$, and somehow also to
the use they made of a second-order polynomial fit. 
Chin and
Krotscheck~\cite{chin} obtained 
$\sigma=0.284$ K\AA$^{-2}$ in an Euler-Lagrange optimized HNC calculation
of $^4$He clusters. Density functional theory including non-local terms are
predictive for the surface tension;~\cite{dalfovo} the different functionals in the
literature~\cite{szyb} show results for the surface tension in the range 0.272-0.287
K\AA$^{-2}$.

\subsection{Density profile}

\begin{figure}
\centering
        \includegraphics[width=0.8\linewidth]{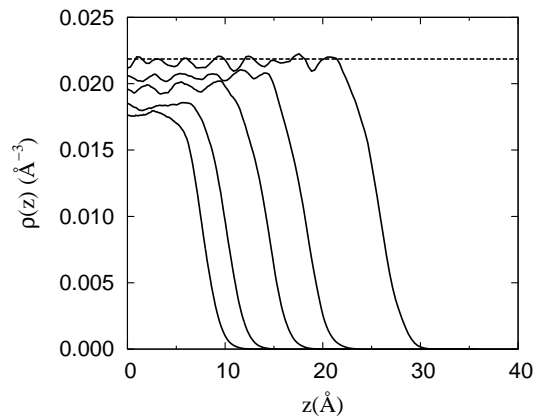}%
	\caption{Density profiles for an increasing number of particles in
	the slab. From left to right, they correspond to $N=$80, 108, 162, 216,
	and 324 atoms. The dashed line shows the equilibrium density of the
	homogeneous liquid, $\rho_0=0.02186$ \AA$^{-3}$.}
\end{figure}

Results for the density profiles of the slab, with different number of atoms
in the simulation cell, are shown in Fig. 2. In order to eliminate any
residual bias, due to the importance sampling trial wave function, pure
estimators have been used.~\cite{pures} When $N$ increases,  the surface
moves along $z$  and the density in the central part increases
progressively towards the equilibrium density of bulk $^4$He
($\rho_0=0.02186$ \AA$^{-3}$). This density is nearly reached for our
largest slab corresponding to $N=324$ atoms. The oscillation in density
which appears in the inner part of the slab does not have any physical meaning
since its amplitude is compatible with the statistical noise there. The
existence of stable oscillations in the surface has been discussed in
several works after the first proposal by Regge.~\cite{regge} Within the present
accuracy, our results exclude this possibility. Nevertheless, the 
results contained in Fig. 2 show a slight shoulder in the inner part of the surface (hardly
observable for the smallest slabs) that could
resemble a Regge oscillation. This property of the $^4$He density
profile has been observed previously in semi-infinite matter and
clusters.~\cite{dalfovo,chin}  

\begin{figure}
\centering
        \includegraphics[width=0.8\linewidth]{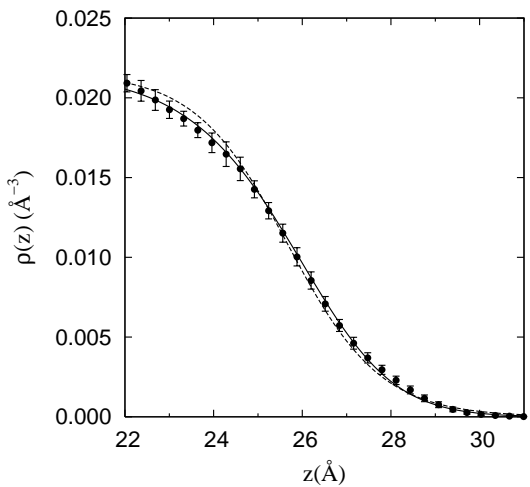}%
	\caption{Symetric vs. asymetric fit to the $^4$He
	density profile. Points with errorbars are the computed values for
	the slab with $N=324$. The dashed  and solid lines correspond to a
	symetric  ($\delta=1$ in Eq. (\protect\ref{rhofit})) and an asymetric
	model ($\delta \neq 1$), respectively.}
\end{figure}

The different slope of the density profile in the inner and outer parts of
the surface has been observed and discussed in previous works. This feature
is directly related to the possible asymmetry of
$\rho(z)$.~\cite{toennies,barranco,lurio2} 
In order to guide our analysis on this point, and
trying to be quantitatively accurate, we have fitted to the DMC data
functions of the type
\begin{equation}
\rho(z)= \frac{\rho_0}{\left( 1 + e^{\beta(z-z_0)} \right)^\delta} \ .
\label{rhofit}
\end{equation}
If $\delta=1$ in Eq. (\ref{rhofit}) the density profile is symmetric;
otherwise, the model is asymmetric with a degree of asymmetry governed by
$\delta$. Using $\chi^2$ as a goodness parameter, we have observed that
the DMC data for the larger slabs ($N \ge 162$) is best reproduced by an
asymmetric fit. On the contrary, for the smaller and less realistic slabs,
the quality of the fit is not improved if $\delta \ne 1$. In Fig. 3,
symmetric and asymmetric fits to the DMC data on the slab with $N=324$ are
shown on top of the simulation results. The differences between the two fits
are certainly small but the asymmetric model matches better the DMC
results, especially in the inner part of the surface where the slope of the profile
is less pronounced. An additional
argument of consistency towards the assessment of an asymmetric profile is
the stability in the value of $\delta$ which is the same for the
three larger slabs, $\delta=1.91(15)$.  

\begin{table}[t]
\centering
\begin{ruledtabular}
\begin{tabular}{cccc} 
$N$ &  $\rho_0$ (\AA$^{-3}$)    & $W$ (\AA)  &  $z_{1/2}$ (\AA)  \\
\hline
80  & 0.0177(2)   &  3.66(5)    &   7.75(5) \\
108 & 0.0184(2)  &  4.01(5)    &  10.09(5)  \\
162 & 0.0198(2)   &  4.57(5)    &  14.19(5) \\
216 & 0.0206(2)   &  4.76(5)    &  18.06(5) \\
324 & 0.0217(2)   &  5.07(5)    &  25.74(5)   \\
\end{tabular}
\end{ruledtabular}
\caption{Central density ($\rho_0$), surface width ($W$), and \textit{size}
of the slab measured with $z_{1/2}$, as a function of the number of atoms
$N$.}
\end{table}

One of the most relevant properties characterizing the surface is its
width,
usually measured as the length ($W$) over which the density decreases from 90 to
10 \% of the inner constant value. Theoretical calculations of the $^4$He
free surface at zero temperature using microscopic
theory~\cite{gernoth,chin,pieper,valles,ceper} or density
functional approaches~\cite{dalfovo,guirao,szyb} predict values in the range 5-7 \AA. These results
are  in overall
agreement with the most recent experimental data at $T=0.45$ K, which point
to a width of 6.5 \AA\ for thick (125 \AA) films.~\cite{penanen} Our results for the width
are reported in Table II as a function of the number of particles in the slab
$N$; the \textit{size} of each slab is given in terms of $z_{1/2}$, defined
as the value of $z$ in which the density is half the density in the inner
part $\rho_0$. The results contained in the Table have been extracted from
the DMC density profile by performing a fit with the asymmetric form given by Eq.
(\ref{rhofit}). If the fit is symmetric ($\delta=1$), $W$ is reduced with
respect to the values reported in the Table but the differences are only
$\sim 0.1$ \AA\ since the degree of asymmetry is rather small. 

\begin{figure}
\centering
        \includegraphics[width=0.8\linewidth]{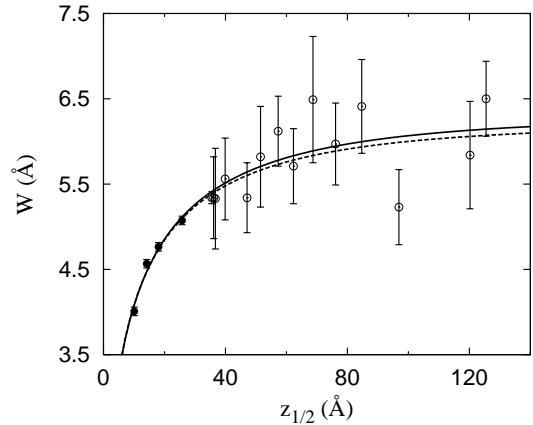}%
	\caption{Interfacial width $W$ as a function of the size of the
	slab, measured in terms of $z_{1/2}$ (solid circles). The open
	circles correspond to experimental data from Penanen \textit{et
	al.}.~\protect\cite{penanen} The solid line is the fit (\ref{winfinit}) considering only our
	theoretical data and the dashed line corresponds to the same
	analytical form but including also the experimental points.} 
\end{figure}

The main
prediction of the present study would be the value of $W$ corresponding to
a free surface, with a central density equal to the bulk equilibrium value
$\rho_0^{\text{bulk}}=0.02186$ \AA$^{-3}$. However, the values reported in
Table II show a dependence on the size of the slab, even for the largest
ones, which makes difficult that estimation. In order to shed light on this
dependence, and trying to analyze possible extrapolations to larger slabs,
the results of the width $W$ are shown as a function of $z_{1/2}$ in Fig.
4. After testing different analytical forms, we have arrived to a rather
simple model that describes quite accurately the DMC data,
\begin{equation}
W(z) = W_\infty \left( 1 - e^{- b \sqrt{z}} \right)  \ ,
\label{winfinit}
\end{equation}
$W_\infty$ being the predicted width of the free surface. The value
obtained is $W_\infty=6.3(4)$ \AA. In Fig. 4, experimental results from
Penanen \textit{et al.},~\cite{penanen}  corresponding to thick $^4$He films adsorbed on a
solid substrate at $T=0.45$ K, are also shown. If the DMC and
experimental data are fitted altogether, according to the empirical law
(\ref{winfinit}), the interfacial width of the free surface becomes     
$W_\infty=6.2(1)$ \AA\ which is compatible with the extrapolated value
considering only the theoretical results. Both extrapolations are in
agreement with the width measured directly for the thickest film of 125
\AA, $W=6.5 \pm 0.5$ \AA.

\begin{figure}
\centering
        \includegraphics[width=0.8\linewidth]{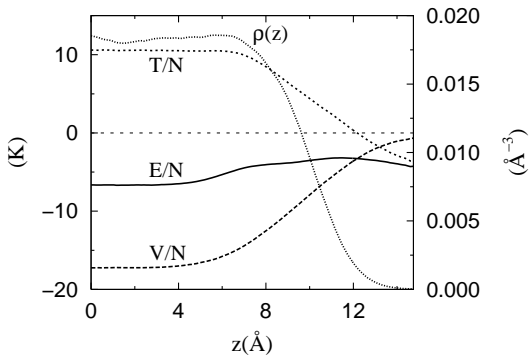}%
	\caption{$z$-dependence of the total ($E/N$), kinetic ($T/N$), and
	potential ($V/N$) energies in the slab geometry. The behavior of
	the energies is compared with the density profile (right scale).
	The data corresponds to a calculation with $N=108$ atoms.}
\end{figure}

The transition from a constant density in the central part of the slab to
a zero density in the outer part of the surface implies a $z$-dependence in
other magnitudes, in particular in the total and partial energies. In Fig.
5, this $z$ dependence is shown in comparison with $\rho(z)$ for a $N=108$
slab. The three functions $E/N(z)$, $V/N(z)$, and $T/N(z)$ are calculated
by averaging the energies of particles located between $z$ and $z+\Delta z$
according to a uniform grid with $\Delta z \simeq 0.1$ \AA\
and, therefore, they are not exact estimations. Nevertheless, the residual
bias coming from the trial wave function is expected to be small and the
qualitative behavior would not change from the one reported in Fig. 5. As
one can see, the potential energy evolves from a constant value in the
inner part (corresponding to the bulk value at the central density of this
slab) to zero when $\rho(z) \rightarrow 0$. Also the total and kinetic
energies in the central part of the slab correspond to the bulk values but,
approaching the surface, the total and kinetic energies reach the same
constant value which is proportional to the analytic
expression~\cite{mantz,belic}
\begin{equation}
E/N (z)= T/N (z) \simeq \lim_{z
\rightarrow  \infty} -\frac{\hbar^2}{2m} \, \frac{\nabla^2
\sqrt{\rho(z)}}{\sqrt{\rho(z)}}    \ .
\label{kinrho}
\end{equation}

\subsection{Distribution and structure functions}

\begin{figure}
\centering
        \includegraphics[width=0.8\linewidth]{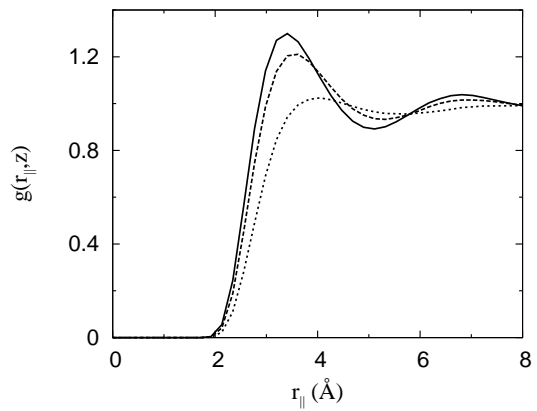}%
	\caption{Two-body radial distribution function for different values
	of $z$, $g(r_{||},z)$. Going from  bottom to top in the height of the
	main peak the data corresponds to $z=$ 11,
	9.5, and 2 \AA\ 	for a slab with 324 atoms.} 
\end{figure}

The slab geometry provides the opportunity of studying the density
dependence of the spatial structure in a quantum liquid  from bulk
to zero densities.  The $z$-dependence of the structure functions is
analyzed by performing slices of variable width $\Delta z$ (larger in the
center of the slab and smaller in the surface) in which the density can be
considered like a constant. The structure of the fluid is then calculated as a
function of $z$, accumulating data corresponding to atoms localized between $z$
and $z+\Delta z$. Following this scheme, we have calculated the two-body radial
distribution function $g(r_{||},z)$ and the results obtained, for a selected
set of $z$ values, are reported in Fig. 6. The functions $g(r_{||},z)$ follow the
evolution of the density: by  increasing $z$  (decreasing $\rho$) the height of
the main peak decreases and moves to larger $r_{||}$ values, reaching in the outer
part of the surface  a very low density regime where $g(r_{||},z)$ resembles the
typical result for a dilute gas.

\begin{figure}
\centering
        \includegraphics[width=0.8\linewidth]{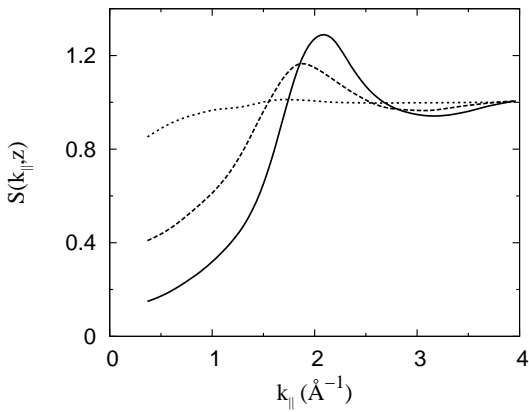}%
	\caption{Static structure function for different values
	of $z$, $S(k_{||},z)$. Going from  bottom to top in the height of the
	main peak the data corresponds to $z=$ 11,
	9.5, and 2 \AA\  for a slab with 324 atoms.} 
\end{figure}

Additional information on the structure of the system, and on its excitation
modes, are contained in the $z$-dependent static structure factor. In its
general form it is defined as
\begin{equation}
S(k_{||},z,z^\prime)= \frac{1}{\left[ N(z) \, N(z^\prime) \right]^{1/2}} \ \left\langle
\rho_{\bm{k}_{||}}(z) \, \rho_{-\bm{k}_{||}}(z^\prime) \right\rangle   \ ,
\label{skz}
\end{equation} 
$\rho_{\bm{k}_{||}}(z)= \sum_{i=1}^{N(z)} e^{i \bm{k}_{||} \cdot \bm{r}_i } $
being the fluctuation density operator for all the particles $N(z)$ in the
slice $(z,z+\Delta z)$, and $\bm{k}_{||}$ a momentum in the $x-y$ plane. Our
calculations have addressed only the diagonal part of this function,
$z= z^\prime$, which in a simplified notation are called $S(k_{||},z)$ in
this work.
Results for $S(k_{||},z)$, corresponding to the same 
slices chosen for $g(r_{||},z)$ in Fig. 6, are shown in Fig. 7. The evolution of
$S(k_{||},z)$ with $z$ follows the same trends observed in Fig. 6 for
$g(r_{||},z)$,
i.e., a behavior essentially determined by the local density in each slice
$\rho(z)$. The behavior when $k_{||} \rightarrow 0$ is known to be different in
the central part and in the surface of the slab. In the central part, where the
mean density is constant, the static structure factor must show a linear
behavior with $k_{||}$, corresponding to the phonon mode of the homogeneous
liquid. On the
contrary, in the surface $S(k_{||},z)$ diverges as $1/\sqrt{k_{||}}$ due to
the presence of ripplons.~\cite{treiner} 
The DMC results shown in Fig. 7 do not show 
signatures of ripplons due probably to restrictions imposed on the lowest
$k_{||}$
value accessible with a finite number of particles.  Recent results for the
structure function,~\cite{ceper} using path integral Monte Carlo (PIMC) with a number of
particles larger than the one used in this work, show a change of behavior
for the smallest $k_{||}$ values and for the most external slices that could be
compatible with the singular behavior induced by ripplons. Notwithstanding,
the size of this signal observed in the PIMC results is appreciably reduced
with respect to a VMC calculation, using shadow wave functions, which
emphasized the importance of ripplons for a correct description of the
$^4$He surface.~\cite{shadow}

\subsection{Condensate fraction}

\begin{figure}
\centering
        \includegraphics[width=0.8\linewidth]{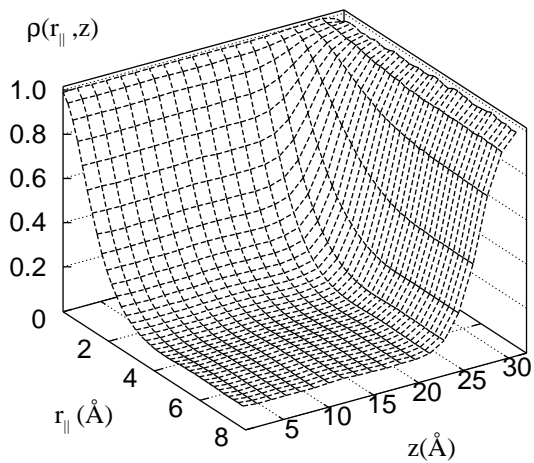}%
	\caption{ One-body density matrix in a $^4$He slab with $N=324$.}
\end{figure}

The presence of off-diagonal long-range order (ODLRO) in a Bose liquid is one of
its more essential properties.~\cite{campbell} The measure of ODLRO in the system is
quantified by the Bose-Einstein condensate fraction $n_0$, i.e., the
fraction of particles occupying the zero-momentum state. This information
is contained in the one-body density matrix
\begin{eqnarray}
\lefteqn{\rho(\bm{r}_1,\bm{r}_1^\prime)=}   \label{onebody} \\
& & N \int d \bm{r}_2 \ldots \bm{r}_N \
\Psi(\bm{r}_1,\bm{r}_2,\ldots,\bm{r}_N)
\Psi(\bm{r}_1^\prime,\bm{r}_2,\ldots,\bm{r}_N) \ .  \nonumber
\end{eqnarray}
The geometry of the slab makes possible to express the one-body density
matrix as a function of $r_{||}$, $z$, and $z^\prime$,
$\rho(r_{||},z,z^\prime)$. Restricting the analysis to the particular case
$z=z^\prime$, the long range behavior of $\rho(r_{||},z)$ gives the local
condensate fraction $n_0(z)$,
\begin{equation}
n_0(z)= \lim_{r_{||} \rightarrow \infty} \frac{\rho(r_{||},z)}{\rho(0,z)} 
\ .
\label{n0local}
\end{equation}

The function $\rho(r_{||},z)$ is sampled in DMC following the standard
procedure, i.e. through the expected value
\begin{equation}
\left\langle \frac{\psi(\bm{r}_1^\prime,\bm{r}_2,\ldots,\bm{r}_N)}
{\psi(\bm{r}_1,\bm{r}_2,\ldots,\bm{r}_N)} \right\rangle  \ ,
\label{rhodmc}
\end{equation}  
accumulating data for a set of slices in the $z$ direction as explained in
the previous subsection. Results for $\rho(r_{||},z)$, corresponding to the
largest slab ($N=324$), are shown in Fig. 8. In this 3D picture, one can see
the constant regime achieved for any $z$ value when $r_{||} \rightarrow \infty$,
which corresponds to the condensate fraction $n_0(z)$ (\ref{n0local}).
Following the evolution of $n_0(z)$ with increasing $z$, one observes a
large enhancement approaching the surface up to values close to 1 which
would correspond to a fully Bose-condensed state. 

\begin{figure}
\centering
        \includegraphics[width=0.8\linewidth]{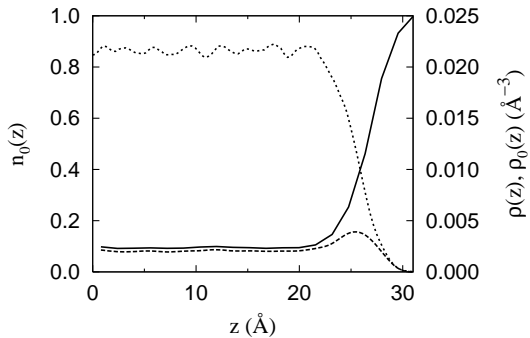}%
	\caption{ $z$ dependence of the condensate fraction in a $^4$He
	slab with $N=324$ (solid line, left scale). The	condensate density 
	$\rho_0(z)$ (long-dashed line) and density  profile $\rho(z)$ 
	(short-dashed line) are also shown (right scale).}
\end{figure}

The change in the condensate fraction as the surface is approached is more
clearly shown in Fig. 9. In the inner part of the slab, where the mean density
is very close to the bulk equilibrium density, $n_0(z)$ is constant and
equal to 0.095(5). When the density starts to decrease, $n_0(z)$ increases
up to almost 1. This result is not surprising since 
 it is well known from bulk calculations that $n_0$ decreases monotonically with the
density. Possible density fluctuations in the surface, due to the presence
of ripplons (surface excited modes), 
that can suppress this nearly Bose-condensate state are not
present in our calculations because the goal of the present work is 
the ground state of the  free $^4$He surface. At
low temperatures, the influence of ripplons in $n_0(z)$ has been studied by
Draeger and Ceperley using the PIMC method;~\cite{ceper} their results show an at most
20 \% reduction of $n_0$, with effects only at extremely low density where,
on the other hand, the statistical signal is rather poor. Figure 9 also
contains the evolution with $z$ of the condensate density, $\rho_0(z)=
n_0(z) \rho(z)$, which shows a clear maximum at approximately $z_{1/2}$ and
then becomes equal to the total density when $\rho(z) \rightarrow 0$. The
resulting picture is in an overall agreement with the VMC results of Lewart
and Pandharipande~\cite{lewart} and with recent DMC studies of trapped hard-core bosons 
 by DuBois and Glyde.~\cite{dubois}

\begin{figure}
\centering
        \includegraphics[width=0.8\linewidth]{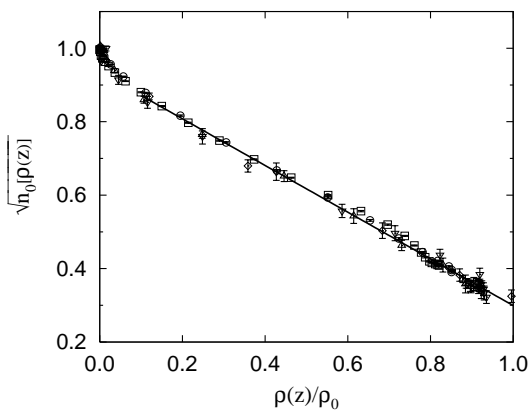}%
	\caption{Condensate amplitude $\sqrt{n_0[\rho(z)]}$ as a function of
	the density $\rho(z)/\rho_0$, with $\rho_0$ the bulk equilibrium
	density. The points correspond to estimations of the condensate
	fraction for all the slabs studied. The solid line corresponds to
	the analytical fit (\protect\ref{pandha_fit})  to data in the density range
	$0.1 \leq \rho(z)/\rho_0 \leq 1$.}
\end{figure}

Collecting data for $n_0$ as a function of the density, and for the
different slabs calculated, one can try to elucidate if they can be well
reproduced by a single analytical form. This was already analyzed by Lewart
\textit{et al.}~\cite{lewart} in a VMC calculation of natural orbitals in $^4$He
clusters. They observed that the condensate fraction for all the clusters
studied could be rather well described by the law
\begin{equation}
\sqrt{n_0[\rho(z)]} = A - B \, \frac{\rho(z)}{\rho_0} \ .
\label{pandha_fit}
\end{equation}
The function $\sqrt{n_0[\rho(z)]}$ can be seen as a condensate amplitude
assuming that $\rho_0(z) \simeq |\Psi_0(z)|^2$, with  $\Psi_0(z)$ the
condensate wave function. The present results for $\sqrt{n_0[\rho(z)]}$,
including the data for all the slabs, are reported in Fig. 10 as a function
of $\rho(z)/\rho_0$, $\rho_0$ being the equilibrium density of bulk
$^4$He. In the range $0.1 \leq \rho(z)/\rho_0 \leq 1$ all the data shows
the same 
linear behavior that can be very well fitted by the empirical law
(\ref{pandha_fit}) with values $A=0.935$ and $B=0.636$. As can be seen in
Fig. 10, the DMC data departs from this linear behavior at very low
densities $ \rho(z)/\rho_0 < 0.1$, approaching faster to the zero value at
zero density.

\section{Discussion}
Using a slab geometry we have presented in this work a DMC study of the
ground-state properties of the free $^4$He surface. The interatomic
potential is highly accurate, and  in the past has allowed for an
excellent reproduction of the experimental equation of state in the bulk
phase.~\cite{borobook} We believe that the combination of an essentially  exact method  and
this very well know interaction between helium atoms guarantees the
accuracy of a microscopic description of the kind intended here. At this
point it is interesting to compare the main results obtained with available
experimental data.

The value of the surface tension of superfluid $^4$He at zero temperature
is one of the main results. Our prediction, $\sigma=0.281(3)$ K\AA$^{-2}$, 
must be compared with the two available and different experimental
measures, $\sigma=0.257 \pm 0.001$ K\AA$^{-2}$ and $\sigma=0.274 \pm 0.002$
K\AA$^{-2}$.~\cite{iino,roche} 
Both experimental determinations correspond to zero temperature extrapolations
using measured values for $\sigma(T)$ and  the Atkins ripplon law for $\sigma(T)$
at very low temperatures.~\cite{edwards}
Our
theoretical result is larger than both measures but it is nearly
statistically compatible with the value  $\sigma=0.274$ K\AA$^{-2}$
obtained by Roche \textit{et al.}.~\cite{roche} It is worth noticing that most of
theoretical estimations of $\sigma$ agree with this larger experimental
value.

A second result which can be compared with experiment is the surface width
$W$. Obviously, the most relevant result would be the \textit{real} width of
the surface, i.e., the asymptotic width for thick slabs. In our results,
$W$ shows a dependence with the size of the slab and, even for the largest
one, this asymptotic regime is not achieved. For the $N=324$ slab the width
is 5.07(5) \AA. Using a reasonable fit to the data we can
estimate the asymptotic surface width, the value obtained being
$W_\infty=6.3(4)$ \AA\
which is in good agreement with a recent experimental measure at
$T=0.44$ K for a 125 \AA\ film, $W=6.5 \pm 0.5$ \AA.~\cite{penanen} On the other hand, the
density profiles, specially the ones with a greater number of particles,
show a slight asymmetry with a deeper slope in the outer part of the
surface. This possible asymmetry, observed also in other theoretical
approaches,~\cite{dalfovo,chin} was analyzed in an x-ray specular reflectivity experiment by
Lurio \textit{et al.}~\cite{lurio2} but uncertainties in the 
scattering amplitude made impossible a clean answer to this point.

Last but not least there is the question of the enhancement of the
condensate fraction in the surface. The experimental measure of $n_0$ in
liquid $^4$He has been elusive for long time due to its small
value.~\cite{glydebook}
Nowadays, deep inelastic neutron scattering has largely improved both its
efficiency and data analysis and therefore there is much more confidence
on the results obtained. Recently, Pearce \textit{et al.}~\cite{pearce} have carried out
neutron scattering experiments on liquid $^4$He adsorbed in thick layers on
an MgO substrate. The analysis of the dynamic structure function obtained
for different number of layers shows unambiguously an increase of the
Bose condensate when the number of layers is reduced. The data obtained are
somehow not completely accurate at the quantitative level but the main
conclusion of the experiment, i.e., the enhancement of $n_0$ in the surface,
is in agreement with our (and others~\cite{lewart,ceper}) theoretical
calculations. It is worth
mentioning that signatures of a large condensate fraction have been also 
observed by Wyatt~\cite{wyatt} in quantum evaporation experiments.

\acknowledgments
We acknowledge financial support from DGI (Spain) Grant No. BFM2002-00466 
and Generalitat de Catalunya Grant No. 2001SGR-00222.

\end{document}